\documentclass[twocolumn]{revtex4-1}

\usepackage{amsmath}
\usepackage{amsfonts}
\usepackage{graphicx}

\begin{document}

\title{Mathematical Melody in Quantum Anomaly via the Path Integral Approach: ~~~~~~~~~~~~~~ A Lesson from the Transverse Current Anomalies in QED}

\author{Israel Weimin Sun}
\email{sunwm@nju.edu.cn}

\affiliation{School of Physics, Nanjing University, Nanjing~210093, the People's Republic of China}

\date{\today}
% Main-Line: from LOGIC to TRUTH. STYLE: Pure mathematical....Highlight: Physical GUAN-NIAN.   English Writings: ZiRan Lian-Cheng Yi-duan-wen.
\begin{abstract}
I address and solve the natural problem of calculating the transverse current anomalies in quantum electrodynamics by means of the path-integral method.
An explicitly divergent and regulator-dependent anomaly term is produced for the vector current, in apparent contradiction with the null-result prediction of
the one-loop perturbative evaluation. This paradox is carefully explained using the concept of infinite-dimensional Grassmann functional integration, signifying
a modification to the conventional wisdom of understanding anomalies in field theory.

\end{abstract}

\maketitle

{\it Introduction.} The quantum anomaly is a basic prediction of the physical world of Quantum Field Theory (QFT). Among all such candidates in gauge theory a simplest but most eminent one should be the Adler-Bell-Jackiw anomaly (ABJ anomaly), or the so-called $U(1)_A$ chiral anomaly, while another natural one is the Weyl anomaly. In the early years of QFT, the quantum anomalies were investigated using the operator methods at the perturbative level \cite{Book-Adler-Jackiw}. In 1979 Kazuo Fujikawa invented an elegant formalism \cite{Fujikawa-PRL} in which the ABJ anomaly is identified with a consequence of the non-invariance of the fermionic functional measure under the chiral rotations. It is a common wisdom in the community that Fujikawa's formulation can now cover all the known quantum anomalies \cite{Fujikawa-book}. In the 2000s a new kind of "quantum anomaly", the so-called "transverse anomaly", has been noticed in the context of the so-called transverse Ward-Takahashi identities. As implied by its very name, a transverse anomaly represents the quantum modification of the classical curl equation (instead of the classical continuity equation) of some symmetry current. In 2001 H.X. He \cite{He-PLB} claimed to have found a transverse axial-vector current anomaly in spinor QED using the gauge-invariant point-splitting technique. Subsequently, Sun ${\it et~al.}$ \cite{Sun} checked his finding by means of an explicit one-loop calculation and made it clear to the community that no transverse axial-vector/vector anomalies exist up to that order.
Incidentally, the apparent "anoamly term" discovered by He is actually a fake one because a careful analysis of its structure reveals that it vanishes. Needless to say, it is advisable to check all these things using Fujikawa's path integral approach. This job was done by several Chinese
authors. In two successive articles \cite{Pseudo-article-1,Pseudo-article-2} they reported a pair of calculations of the transverse axial-vector/vector anomaly based on Fujikawa's original idea and claimed to have found a novel (transverse) anomaly term, in disagreement with the result of Sun ${\it et~al.}$. Unfortunately, a simple glimpse of their "derivations" reveals that those authors had made a sequence of dirty and inexcusable mistakes which puts their whole conclusion in jeopardy. Because of this sad situation, in this Letter I shall report my own check of this issue using the path integral approach.
Contrary to those authors' findings, I show that the standard technology {\it $\grave{a}$ la} Fujikawa could no longer produce a definite and finite "transverse anomaly" for the vector current  (although it seems to imply a trivial null-anomaly term for the case of axial-vector current). I argue that this strange situation actually implies  an inevitable breakdown of the normal change-of-variable rule of Grassmann integration since in such a case the "infinite-dimensional functional Jacobian" involved in the relevant field transformations is undoubtedly too poorly behaved to be mathematically well defined. In such a circumstance we have to say that the standard Fujikawa's "philosophy" ceases to provide a universal framework, or a pure Paradise, to account for all the physical quantum anomalies in the real world.

{\it A global-type analysis of the whole issue---from logic to truth.} As a first step, let me briefly display the basic logic of the whole process. One starts from the Euclidean generating functional of a quadratic fermionic system coupled to an external $U(1)$ gauge potential
\begin{eqnarray}
\nonumber Z[A] &=& \int {\mathcal D}{\bar \psi}{\mathcal D}\psi~e^{-S[{\bar \psi},\psi;A]} \\
\nonumber &=& \int {\mathcal D}{\bar \psi}{\mathcal D}\psi ~e^{-\int d^4x{\bar \psi}({\not\! D}+m)\psi} \\
          &=& \int {\mathcal D}{\bar \psi}{\mathcal D}\psi ~e^{-\int d^4x\frac{1}{2}({\bar \psi}(\overrightarrow{{\not\! D}}+m)\psi-{\bar \psi}(\overleftarrow{{\not\! D}}-m)\psi)},
\end{eqnarray}
where the covariant derivative has two forms: $\overrightarrow{D}_\mu=\overrightarrow{\partial}_\mu-ie A_\mu$,$\overleftarrow{D}_\mu=\overleftarrow{\partial}_\mu+ie A_\mu$, respectively. The four hermitian matrices $\gamma_\mu~(\mu=1,2,3,4)$~obey the Clifford algebra relation $\{\gamma_\mu,\gamma_\nu\}=2\delta_{\mu\nu}$.
Subsequently, one performs the following change-of-variable of the Grassmann spinor field
\begin{eqnarray}\label{Grassmann-pattern}
\nonumber \psi'(x)&=& e^{i\Omega \alpha(x)}\psi(x) \\
{\bar \psi'}(x)&=& {\bar \psi}(x)e^{i{\tilde \Omega} \alpha(x)}
\end{eqnarray}
to generate a sequence of events which are all I want. As it stands alone in (\ref{Grassmann-pattern}), the couple of matrices $(\Omega,{\tilde \Omega})$ could be arbitrarily chosen, since a pair of Grassmann fields $({\bar \psi},\psi)$ is independent integration variables in the Euclidean metric. In principle, each member of the pair $(\Omega,{\tilde \Omega})$ can be independently expanded according to the 16 bases of the Clifford algebra, however, for my personal purpose I
will only consider two special cases: ${\tilde \Omega}=\pm \Omega$ which itself is general enough to accommodate the generation of the possible transverse anomalies of the axial-vector/vector current.

Now, let me proceed further to the path integral formalism itself. As is known to all, the standard chiral anomaly originates from the formal non-invariance of the Grassmann integration measure ${\mathcal D}{\bar \psi}{\mathcal D}\psi$ under the chiral rotation: $\Omega={\tilde \Omega}=\gamma_5$.
Mathematically, one needs to give a precise definition to the functional Jacobian which appears both naturally and inevitably before
the formal object ${\mathcal D}{\bar \psi}{\mathcal D}\psi$. The natural idea is very simple. One should
extract something sensible from a vague formalism by some regularization method, which in fact works perfectly well in the case of chiral anomaly. In the following,
I want to tell all my readers a wonderful (hopefully less depressing) story, showing that the BIBLE of MATHEMATICS would not give any person a piece of gem
for nothing because the Glorious GOD would only be willing to save the Genuine Soul of an Enlightened FAUST!

Such a graceful story begins like this. First, please remember that a routine evaluation of everything produces two pieces as shown
below (assuming $\alpha$ to be an infinitesimal one):
\begin{eqnarray}
\nonumber && S[{\bar \psi},\psi;A] = S[{\bar \psi'},\psi';A]+\Delta S \\
\nonumber &&\Delta S =  \\
\nonumber && \int d^4 x \big(\frac{i}{2}\partial_\rho({\bar \psi'}(x)[\gamma_\rho,\Omega]_\mp \psi'(x)) +\frac{i}{2}\partial_\rho {\bar \psi'}(x)[\gamma_\rho,\Omega]_\pm \psi'(x)\\
\nonumber && ~~~~-\frac{i}{2}{\bar \psi'}(x)[\gamma_\rho,\Omega]_\pm \partial_\rho \psi'(x)-e {\bar \psi'}(x)[\gamma_\rho,\Omega]_\pm \psi'(x) A_\rho(x) \\
&& ~~~~-i m{\bar \psi'}(x)(\Omega\pm \Omega)\psi'(x)\big)\alpha(x),
\end{eqnarray}
\begin{eqnarray}
\nonumber {\mathcal D}{\bar \psi}{\mathcal D}\psi &=& e^{i\int d^4x \alpha(x)\sum_{n}\phi^\dag_n(x)(\Omega+{\tilde \Omega})\phi_n(x)}{\mathcal D}{\bar \psi'}{\mathcal D}\psi' \\
&=& J(A;\alpha){\mathcal D}{\bar \psi'}{\mathcal D}\psi',
\end{eqnarray}
where $\phi_n(x)$ is a sequence of complete orthonormal spinor eigenfunctions of the hermitian Euclidean Dirac operator: $i {\not \! D} \phi_n(x)=\lambda_n \phi_n(x)$.
Then, a direct and naive rearrangement of everything yields a formal identity
\begin{equation}\label{Change-of-variable}
\int {\mathcal D}{\bar \psi}{\mathcal D}\psi e^{-S[{\bar \psi},\psi;A]}=\int {\mathcal D}{\bar \psi}{\mathcal D}\psi~J(A;\alpha)e^{-(S[{\bar \psi},\psi;A]+\Delta S)}
\end{equation}
whose $\mathcal{O}(\alpha)$-order coefficient shows the following pattern
\begin{eqnarray}
\nonumber &&\partial_\rho \langle \frac{1}{2}{\bar \psi}(x)[\gamma_\rho,\Omega]_\mp \psi(x)\rangle_{A}  \\
\nonumber &=& \langle \frac{1}{2}{\bar \psi}(x)[\gamma_\rho,\Omega]_\pm(\overrightarrow{\partial}_\rho-\overleftarrow{\partial}_\rho) \psi(x)\rangle_{A} -ie
\langle {\bar \psi}(x)[\gamma_\rho,\Omega]_\pm \psi(x)\rangle_{A} \\
\nonumber  &&~A_\rho(x)+m \langle {\bar \psi}(x)(\Omega\pm \Omega)\psi(x)\rangle_{A} \\
&&+\sum_n \phi^\dag_n(x)(\Omega+{\tilde \Omega})\phi_n(x),
\end{eqnarray}
which implies a particular set of relations among the various path-integral correlation functions in the external background gauge field.
This is my raw material. To produce something interesting to me, I shall make two choices. In the first choice I take $\Omega=\sigma_{\mu\nu}$ and set ${\tilde \Omega}=\Omega$,
whereas in the second choice I take $\Omega=\gamma_5 \sigma_{\mu\nu}$ and use ${\tilde \Omega}=-\Omega$. In both cases the numerical double-indices $(\mu,\nu)$ are adjusted as I wish. In these two design schemes a direct use of the Euclidean Dirac algebra will give us two kinds of equality. More specifically, in the first case $[\gamma_\rho,\sigma_{\mu\nu}]=2i(\delta_{\rho\nu}\gamma_\mu-\delta_{\rho\mu}\gamma_\nu)$ yields
\begin{eqnarray}\label{VC-Equation}
\nonumber &&\partial_\mu \langle {\bar \psi}(x)\gamma_\nu \psi(x)\rangle_{A}- \partial_\nu \langle {\bar \psi}(x)\gamma_\mu \psi(x)\rangle_{A} \\
\nonumber &=& \langle {\bar \psi}(x)\frac{i}{2}\{\gamma_\rho,\sigma_{\mu\nu}\}(\overrightarrow{\partial}_\rho-\overleftarrow{\partial}_\rho) \psi(x)\rangle_{A} +e
\langle {\bar \psi}(x)\{\gamma_\rho,\sigma_{\mu\nu}\} \psi(x)\rangle_{A} \\
 &&~A_\rho(x)+2m i\langle {\bar \psi}(x)\sigma_{\mu\nu}\psi(x)\rangle_{A}+2i\sum_n \phi^\dag_n(x)\sigma_{\mu\nu}\phi_n(x),
\end{eqnarray}
which is just the vector current curl equation with a potential anomaly term, while in the second case $\{\gamma_\rho,\gamma_5\sigma_{\mu\nu}\}=2i(\delta_{\rho\nu}\gamma_\mu\gamma_5-\delta_{\rho\mu}\gamma_\nu\gamma_5)$ yields
\begin{eqnarray}
\nonumber  &&\partial_\mu \langle {\bar \psi}(x)\gamma_\nu\gamma_5 \psi(x)\rangle_{A}- \partial_\nu \langle {\bar \psi}(x)\gamma_\mu \gamma_5\psi(x)\rangle_{A}\\
\nonumber  &=& \langle {\bar \psi}(x)\frac{i}{2}[\gamma_\rho,\gamma_5\sigma_{\mu\nu}](\overrightarrow{\partial}_\rho-\overleftarrow{\partial}_\rho) \psi(x)\rangle_{A} \\
&&+e\langle {\bar \psi}(x)[\gamma_\rho,\gamma_5\sigma_{\mu\nu}] \psi(x)\rangle_{A}A_\rho(x),
\end{eqnarray}
which is the axial-vector current curl equation with a null anomaly term.

The true tragedy shows up immediately, as will be described in detail below. Following Fujikawa's original idea, I introduce a smooth function $f: [0,\infty) \rightarrow \mathbb{R}$ satisfying the following two conditions:
\begin{eqnarray}
&&({\rm a}): f(0)=1,~\forall~n \geq 1~ f^{(n)}(0)~{\rm exists~and~is~finite}, \\
&&({\rm b}): \lim_{s \rightarrow \infty} s^{m}f^{(n)}(s)=0 ~~ \forall ~m \geq 0 ~~\forall ~n \in \mathbb{Z }^{+},
\end{eqnarray}
as a regulator to damp the large spectrum-point contribution to the otherwise undefined pointwise-convergence-type summation in the RHS of Eq. (\ref{VC-Equation})
so that a finite outcome could be produced.

Condition (a) is elementary, whereas Condition (b) implies that all the derivatives $f^{(n)}(s)$ are fast-decreasing as $s \rightarrow \infty$. As to the summation itself, I would like to warn the readers that the discrete eigenvalues $\{\lambda_n\}$, or in mathematical jargon, the point spectrum of the Dirac operator $i{\not \! D}$, are reflection invariant w.r.t. the isolated point $\lambda=0$ because of the sign reversal of the eigenvalues $\lambda_n$ under the discrete $\mathbb{Z}_2$ map $\phi_n \mapsto \gamma_5 \phi_n$, and consequently the summation is actually a two-sided one.

Such a regulator $f(s)$ could be many, for example, a Gaussian cutoff $f(\frac{\lambda_n^2}{M^2})=e^{-\lambda_n^2/M^2}$ as introduced by Fujikawa himself will do the job,
although many more alternatives could be adopted in a natural way. When one deals with the chiral anomaly in spinor QED, all these choices are equivalent and
effectively play the same role known as the gauge-invariant mode cutoff regularization \cite{Fujikawa-book}. Unfortunately, in the circumstance explored in this article
this regulator-independence is cruelly lost to a large extent, which makes an honest man unwilling to accept the Minimal Physical Philosophy Doctrine: Perfect Logic
will undoubtedly Lead to Absolute Truth. This situation is not artificial, but could only be true. OK. The logical part is over, let me turn to the Truth.

As explained before, my strategy is standard, just like how Kazuo Fujikawa deals with his cherished chiral anomaly. I try to
give a formal definition to the "anomaly term" part:
\begin{eqnarray}
\nonumber &&\lim_{M \rightarrow \infty} \sum_{n} f(\frac{\lambda_n^2}{M^2})\phi^\dag_n(x)\sigma_{\mu\nu}\phi_n(x) \\
\nonumber &=& \lim_{M \rightarrow \infty} \sum_{n} \phi^\dag_n(x)\sigma_{\mu\nu}f(\frac{(i {\not \! D})^2}{M^2})\phi_n(x)\\
 &=& \lim_{M \rightarrow \infty}\int \frac{d^4 k}{(2\pi)^4}tr \big[ \sigma_{\mu\nu}e^{ik\cdot x} f(\frac{(i {\not \! D})^2}{M^2})e^{-ik\cdot x} \big],
\end{eqnarray}
where the trace operation acts on the Dirac indices. Then, note that an operator identity $gf(C)g^{-1}=f(gCg^{-1})$ implies $e^{ik\cdot x} f(\frac{(i {\not \! D})^2}{M^2})e^{-ik\cdot x}=f(\frac{-{\not \! D}^2+k^2+2i k_\mu D_\mu}{M^2})$, which contains implicit evaluation on the constant function unity. Thus, after a further
rescaling $k \rightarrow M k$ the anomaly term can be written as
\begin{equation}
\lim_{M \rightarrow \infty}M^4 \sum_{n=0}^{\infty}\frac{1}{n!}\int \frac{d^4 k}{(2\pi)^4}f^{(n)}(k^2)tr \big[ \sigma_{\mu\nu}\big( \frac{2i k_\mu D_\mu}{M}-\frac{{\not \! D}^2}{M^2}\big)^n \big].
\end{equation}
The task that remains is to find the final answer. I shall do this term by term.

First, as it stands, the $n=0$ term vanishes trivially, since $tr \sigma_{\mu\nu}=0$. The $n=1$ term reads
\begin{eqnarray}
\nonumber && M^4 \int \frac{d^4 k}{(2\pi)^4}f'(k^2)~tr \big[ \sigma_{\mu\nu}\big( \frac{2i k_\mu D_\mu}{M}-\frac{{\not \! D}^2}{M^2}\big) \big] \\
&=& M^2 \int \frac{d^4 k}{(2\pi)^4}f'(k^2)~tr [\sigma_{\mu\nu}(-{\not \! D}^2)],
\end{eqnarray}
where one finds $\int \frac{d^4 k}{(2\pi)^4}f'(k^2)=-\frac{1}{16 \pi^2}\int_{0}^{\infty} ds f(s)$ and $tr [\sigma_{\mu\nu}(-{\not \! D}^2)]=-4e F_{\mu\nu}$. Hence, the net contribution is $\frac{M^2}{4\pi^2}\int_{0}^{\infty} ds f(s)e F_{\mu\nu}$. Then, one examines the $n=2$ term
\begin{eqnarray}
\nonumber && M^4\frac{1}{2!} \int \frac{d^4 k}{(2\pi)^4}f''(k^2)tr \big[ \sigma_{\mu\nu}\big( \frac{2i k_\mu D_\mu}{M}-\frac{{\not \! D}^2}{M^2}\big)^2 \big] \\
   &=& \frac{1}{2!} \int \frac{d^4 k}{(2\pi)^4}f''(k^2)tr [ \sigma_{\mu\nu}{\not \! D}^4 ].
\end{eqnarray}
A simple calculation reveals
\begin{equation}
\int \frac{d^4 k}{(2\pi)^4} f''(k^2)=-\frac{1}{16 \pi^2}\int_{0}^{\infty} ds f'(s)=\frac{1}{16 \pi^2},
\end{equation}
and $tr [\sigma_{\mu\nu}{\not \! D}^4]$ can be straightforwardly evaluated to be $4e(\Box F_{\mu\nu}-2ie\partial_\rho(A_\rho F_{\mu\nu})+2(ie)^2 F_{\mu\nu}A^2)$. The outcome
is $\frac{e}{8 \pi^2}(\Box F_{\mu\nu}-2ie\partial_\rho(A_\rho F_{\mu\nu})+2(ie)^2 F_{\mu\nu}A^2)$. The analysis of $n=3$ term is more involved. Upon expanding
out the whole expression $M^4\frac{1}{3!} \int \frac{d^4 k}{(2\pi)^4}f'''(k^2)tr \big[ \sigma_{\mu\nu}\big( \frac{2i k_\mu D_\mu}{M}-\frac{{\not \! D}^2}{M^2}\big)^3 \big]$
into various parts, all one needs to do is just a simple power-counting. Among all the $2^3=8$ terms, the only member that is potentially divergent, the one of the structure $(\frac{2i k_\mu D_\mu}{M})^3$, actually vanishes since $tr \sigma_{\mu\nu}=0$, while the rest of them have different fates according to their own structures, which I shall spell out now. In fact, in the lattice format $(\frac{1}{M}+\frac{1}{M^2})\otimes(\frac{1}{M}+\frac{1}{M^2})\otimes(\frac{1}{M}+\frac{1}{M^2})$, one could pick out 1 $\frac{1}{M}$ and 2 $\frac{1}{M^2}$, which form 3 terms, all being marked with a power $\mathcal{O}(\frac{1}{M})$, alternatively, one picks out 2 $\frac{1}{M}$ and 1 $\frac{1}{M^2}$, which corresponds to another 3 terms with no powers of $M$. The remaining one is formed by combining 3 $\frac{1}{M^2}$ in such a format, which has a
net power $\mathcal{O}(\frac{1}{M^2})$. Thus, only 3 terms among them survive the $M \rightarrow \infty$ limit:
\begin{eqnarray}
\nonumber &&\frac{1}{3!} \int \frac{d^4 k}{(2\pi)^4}~4 k_\rho k_\sigma f'''(k^2)tr\big[\sigma_{\mu\nu}(D_\rho D_\sigma {\not \! D}^2+D_\rho {\not \! D}^2 D_\sigma+{\not \! D}^2 D_\rho D_\sigma )\big] \\
&&= \frac{1}{3!}\int \frac{d^4 k}{(2\pi)^4}k^2 f'''(k^2)tr\big[\sigma_{\mu\nu}(D^2 {\not \! D}^2+D_\rho {\not \! D}^2 D_\rho+{\not \! D}^2 D^2 )\big],
\end{eqnarray}
where I have used the tensorial substitution rule: $k_\mu k_\nu \rightarrow \frac{1}{n}\delta_{\mu\nu}k^2$ under an $SO(n)$-invariant Euclidean space integration.
The remaining computation is quite simple. The coefficient factor is found to be
\begin{equation}
\int \frac{d^4 k}{(2\pi)^4} k^2 f'''(k^2)=\frac{1}{8 \pi^2}\int_{0}^{\infty} ds  f'(s)=-\frac{1}{8 \pi^2},
\end{equation}
which, together with an easy and elementary evaluation of the core of the trace part, produces a final contribution $-\frac{e}{12 \pi^2}(\Box F_{\mu\nu}-3ie\partial_\rho(A_\rho F_{\mu\nu})+3(ie)^2 F_{\mu\nu}A^2)$.

The estimates of the terms with $n \geq 4$ are pretty similar to our previous work except for some finer points. For instance, the $n=4$
term shows a structure $M^4\frac{1}{4!} \int \frac{d^4 k}{(2\pi)^4}f^{(4)}(k^2)tr \big[ \sigma_{\mu\nu}\big( \frac{2i k_\mu D_\mu}{M}-\frac{{\not \! D}^2}{M^2}\big)^4 \big]$,
which contains a nominal $\mathcal{O}(1)$ term generated from a pure monomial of the type $(\frac{2i k_\mu D_\mu}{M})^4$, all the other terms being of a negative power of $M$ under an easy inspection. However, such an $\mathcal{O}(1)$ term is only superficial and actually vanishes owing to $tr \sigma_{\mu\nu}=0$. All the remaining terms vanish in the
$M \rightarrow \infty$ limit, since each of them produces a convergent 4-momentum integration because of the rapid decrease property of $f^{(4)}(k^2)$ at the infinite boundary
of the Euclidean 4-space.

The general logic for such an argument can be repeated for all the $n \geq 5$ terms. Here, for the sake of mathematical elegance, I would like to provide a more compact proof
by means of the Dominated Convergence Theorem \cite{Rudin}. The idea is pretty simple. For all $n \geq 5$ terms, the expansion of the bulk $M^4 \frac{1}{n!}\int \frac{d^4k}{(2\pi)^4}f^{(n)}(k^2)tr \big[ \sigma_{\mu\nu}\big( \frac{2i k_\mu D_\mu}{M}-\frac{{\not \! D}^2}{M^2}\big)^n \big]$ will produce a finite sequence of terms, everyone of which is of the order $\mathcal{O}(\frac{1}{M^m})$ with some $m \geq 1$. If one can freely interchange the order of the limit process and the Euclidean momentum space integration
operation, one will immediately obtain the expected results. This is definitely so because the composition structure of a generic $n$-th term is of the following pattern:
\begin{eqnarray}
\nonumber &&\int \frac{d^4k}{(2\pi)^4} \sum_{{\rm type}}f^{(n)}(k^2)\times \stackrel{{\rm standard~monomial}}{(k_{\rho_1}k_{\rho_2} \cdots k_{\rho_s})}\\
&& \times ({\rm differential~operator})\mapsto 1 \frac{1}{M^{m_{type}}},
\end{eqnarray}
in which some "standard monomial" of the type $(k_{\rho_1}k_{\rho_2} \cdots k_{\rho_s})$ could be degenerate to $1$ in a trivial manner where appropriate and the associated
differential operator corresponding to that specific $(k_{\rho_1}k_{\rho_2} \cdots k_{\rho_s})$-monomial should be understood to act on the constant function unity, consequently
this integral is dominated by
\begin{eqnarray}\label{DominanceForm}
\nonumber &&\int \frac{d^4k}{(2\pi)^4} \sum_{{\rm type}}|k_{\rho_1}k_{\rho_2} \cdots k_{\rho_s} f^{(n)}(k^2) | \frac{1}{M^{m_{type}}}  \\
&& \times( {\rm some~type~of~body}) ,
\end{eqnarray}
where the so-called "body" refers simply to the norm of some numerical function of the Euclidean space point $x=(x_1,x_2,x_3,x_4)$ which results from the action of some numerical differential operator on the constant function 1, hence the exact integrand of
the total piece (\ref{DominanceForm}) should be dominated by the Master-Form:
\begin{equation}
\sum_{{\rm term~type}}|k_{\rho_1}k_{\rho_2} \cdots k_{\rho_s} f^{(n)}(k^2) |\times( {\rm some~type~of~body}) (x),
\end{equation}
as long as the free regularization parameter $M$ is sufficiently large during the limiting process $M \rightarrow \infty$. The Master-Form is a legitimate one because
for each "term-type" the smooth function $k_{\rho_1}k_{\rho_2} \cdots k_{\rho_s} f^{(n)}(k^2)$ is a member of $L^{1}(\mathbb{R}^4; \frac{d^4k}{(2\pi)^4})$ and there are only
a finite number of such summands for any $n \geq 5$.

Thus, I conclude that such a sequence of contributions terminates at the $n=4$ term. A simple summation of all the nonvanishing contributions leads to a so-called
"transverse anomaly term":
\begin{eqnarray}\label{AnomalyForm}
\lim_{M \rightarrow \infty}\frac{i M^2}{2\pi^2}\int_{0}^{\infty}ds f(s)e F_{\mu\nu}+\frac{ie}{12\pi^2}\Box F_{\mu\nu}.
\end{eqnarray}

It is apparent to see that such an "anomaly term" is an irregular mathematical object which diverges badly as $M \rightarrow \infty$, so anyone is quite perplexed at
its true physical meaning. Needless to say, the very existence or non-existence of "transverse anomaly" should definitely
be checked in all possible ways. As to (\ref{AnomalyForm}) itself, one could only extract a definite physical prediction from its body by setting
$\int_{0}^{\infty}ds f(s)=0$, thus completely forgetting about the annoying $\mathcal{O}(M^2)$ divergence. In principle, such a prescription is permitted, because the regulator $f(s)$ can be rather freely chosen. Unfortunately,
a simple Gaussian regulator $f(s)=e^{-s}$ fails to meet this requirement, and the celebrated "$f(s)$-independence", which marks the true success of Fujikawa's
path-integral treatment of chiral anomaly, is completely lost in such a simple and innocent circumstance. One ought to remember that the one-loop perturbative evaluation of the vector current "transverse anomaly" produces a null result !
Then, which side goes wrong? Is Mathematics unwilling to show us a friendly hand, or, dose Physics make a joke to everyone of us ?

{\it A solution to the paradox: Fujikawa's Path-Integral Dogma needs to go away. }
Here I show my own solution of this puzzle. According to my understanding, a piece of combined interaction between Logic and Truth accounts for everything.
On the logical side, if one admits
the validity of the Fujikawa's method, then the non-uniquely defined form (\ref{AnomalyForm}) needs to correspond exactly to the various calculational results from
all the different regularization modes in the loop-diagram treatment. However, as was pointed out by Sun ${\it et~al.}$ in their paper, both Pauli-Villars-type regularization and the continuous dimensional regularization yield a null-anomaly result at the one-loop level. Then, how does one's innocent mind face such a dreadful scene ? This is just a problem of Truth. In fact, as can be envisaged by everyone's metaphysical thinking or idea, the two processes, namely, the operation of "rotating" the Grassmann integration variables $({\bar \psi,\psi})$ and that of defining the (one-loop-order) correlation function by a particular regularization method, are orthogonal to each other, and the Truth should
be: all the correlators contained in (\ref{VC-Equation}) could no longer be flexibly altered to such an extent as to match the arbitrariness of the mathematically undefined
piece (\ref{AnomalyForm}) !

The genuine mathematical explanation can be phrased as follows. An irregularity of the functional Jacobian factor would only imply the breakdown of the
usual change-of-variable rule for the circumstance of an infinite-dimensional Grassmann functional integration, e.g., the one in (\ref{Change-of-variable}), because any definite mathematical process in the metaphysical sense could only lead to a series of exact identities relating the various path-integral-type correlators as displayed in (\ref{VC-Equation}), while a free mind is willing to write down a piece of logical thought:
\begin{eqnarray}
\nonumber && \int {\mathcal D}{\bar \psi}{\mathcal D}\psi e^{-{\bar \psi}K\psi}\equiv \int {\mathcal D}{\bar \psi}{\mathcal D}\psi(\det M)^{-1}(\det N)^{-1}e^{-{\bar \psi}M K N\psi}  \\
&& \Longleftrightarrow \det K =(\det M)^{-1}\times (\det N)^{-1}\times \det (MKN),
\end{eqnarray}
which implies immediately $(\det M)(\det N)= \det (MKN)/\det K$, the latter being the various moments of the formal "Grassmann-type Gaussian measure" $e^{-{\bar \psi}K\psi}{\mathcal D}({\bar \psi}\psi)$, which in our case are defined loop-wise in terms of the background gauge field and thus could not be identified with an inherently mathematically ambiguous object $(\det M)(\det N)$, hence a contradiction.

Therefore, my global conclusion is as follows. First, the so-called transverse axial-vector/vector current anomaly is definitely absent in QED, and consequently the Path-Integral-type formulation of quantum anomaly as developed by Kazuo Fujikawa would collapse in such a case. The Gentleman of Mathematics has an Innocent Mind
and he is always willing to give you a piece of Golden Dollar for a Full Book.

% The Cloud and the Moon are common guests of the Sky .

I thank Professor Fan Wang who is my PhD supervisor for suggesting me undertake this project and his valuable instructions.
This work is supported in part by the Natural Science Funds of Jiangsu Province of China under Grant No. BK20151376.

\end{document}